\documentclass{PoS}
\usepackage{amssymb,amsmath,epsfig}
\newcommand{\eq}[1]{Eq.~(\ref{#1})}

\newcommand{\be}{\begin{equation}}
\newcommand{\ee}{\end{equation}}
\newcommand{\bea}{\begin{eqnarray}}
\newcommand{\eea}{\end{eqnarray}}
\newcommand{\ben}{\begin{eqnarray*}}
\newcommand{\een}{\end{eqnarray*}}

\newcommand{\DS}{Dyson-Schwinger }
\newcommand{\BS}{Bethe-Salpeter }

\newcommand{\w}{\omega}

\newcommand{\ga}{\gamma}
\newcommand{\G}{\Gamma}
\newcommand{\de}{\delta}

\newcommand{\si}{\sigma}

\newcommand{\ro}{\rho}

\newcommand{\La}{\Lambda}
\newcommand{\ka}{\kappa}

\newcommand{\pd}{\partial}

\renewcommand{\div}{\vec{\nabla}}

\newcommand{\s}[2]{{#1}\!\cdot\!{#2}}
\newcommand{\ov}[1]{\overline{#1}}

\newcommand{\ev}[1]{<\!\!{#1}\!\!>}

\title{Gap and Bethe-Salpeter equations in Coulomb gauge}
\ShortTitle{Gap and Bethe-Salpeter equations in Coulomb gauge}
\author{\speaker{Peter Watson}%
         \thanks{Work supported by the Deutsche Forschungsgemeinschaft 
(DFG) under contracts no. DFG-Re856/6-2,3.}\\
Institut f\"ur Theoretische Physik, Universit\"at T\"ubingen, 
Auf der Morgenstelle 14, D-72076 T\"ubingen, Deutschland\\
        E-mail: \email{watson@tphys.physik.uni-tuebingen.de}}
\author{Hugo Reinhardt\\
Institut f\"ur Theoretische Physik, Universit\"at T\"ubingen, 
Auf der Morgenstelle 14, D-72076 T\"ubingen, Deutschland
}
\abstract{I will discuss the gap and Bethe-Salpeter equations for quantum chromodynamics in Coulomb gauge under a leading order truncation scheme.  Within this scheme, the role of charge conservation and the cancellation of infrared divergences becomes particularly clear.  The quark gap equation exhibits not only chiral symmetry breaking, but explicitly reproduces the heavy quark limit.  The corresponding gluon equation has a massive solution with interesting nonperturbative renormalization properties.  I will further discuss various aspects of the Bethe-Salpeter equation for pseudoscalar and vector mesons with arbitrary quark masses and present numerical results for the meson masses and leptonic decay constants.}
\FullConference{Xth Quark Confinement and the Hadron Spectrum,\\
		October 8-12, 2012\\
		TUM Campus Garching, Munich, Germany}
\begin{document}
\section{Coulomb gauge at leading order}
The first part of this talk concerns the construction of a leading order truncation to the \DS equations of Coulomb gauge QCD \cite{Watson:2011kv}.  Let us begin by considering the following (standard) functional integral
\be
Z=\int{\cal D}\Phi e^{\imath{\cal S}},\;\;\;\;{\cal S}={\cal S}_q+\int dx(E^2-B^2)/2,
\ee
where the action (${\cal S}$) is split into a quark component, ${\cal S}_q$, and the Yang-Mills part.  ${\cal D}\Phi$ generically denotes the integration measure.  The chromomagnetic field, $\vec{B}$, will not concern us in the following.  The chromoelectric field, $\vec{E}$, is given by (superscript indices $a,\ldots$ denote the color index in the adjoint representation)
\be
\vec{E}^a=-\pd_0\vec{A}^a-\vec{D}^{ab}A_0^b,\;\;\;\;\vec{D}^{ab}=\de^{ab}\div-gf^{acb}\vec{A}^c,
\ee
where $\vec{D}$ is the spatial component of the covariant derivative in the adjoint color representation (the $f$ are the usual $SU(N_c)$ structure constants).  We work in Coulomb gauge ($\div\cdot\vec{A}=0$), for which the corresponding Faddeev-Popov (FP) operator is $-\div\cdot\vec{D}$.  There are two important points: the FP operator involves purely spatial operators and the chromoelectric field is linear in the temporal component of the gauge field, $A_0$.  We now convert to the first order formalism by introducing an auxiliary field $\vec{\pi}$ via the identity
\be
\exp{\left\{\imath\int dx\,E^2/2\right\}}=\int{\cal D}\vec{\pi}\exp{\left\{\imath\int dx\left[-\pi^2/2-\vec{\pi}^a\cdot\vec{E}^a\right]\right\}}.
\ee
The field $\vec{\pi}$ is split into transverse ($\div\cdot\vec{\pi}^\perp=0$, henceforth we drop the $\perp$) and longitudinal ($\div\phi$) parts.  Since the action is now linear in $A_0$, we can integrate it out, to give
\be
Z=\!\int\!\!{\cal D}\Phi\de\left(\div\cdot\vec{A}\right) \de\left(\div\cdot\vec{\pi}\right)\mbox{Det}\left[-\div\cdot\vec{D}\right]\de\left(\div\cdot\vec{D}\phi+\rho\right)e^{\imath{\cal S}'},\;\;
\rho^a=gf^{abc}\vec{A}^b\cdot\vec{\pi}^c+g\ov{q}\left[\ga^0T^a\right]q,
\ee
where $\rho$ is the color charge (including the quark component, with quark field $q$ and the Hermitian color generator $T^a$).  The $\phi$-field can also be integrated out to cancel the FP determinant (Coulomb gauge is formally ghost free).  However, noting the temporal zero modes of the FP operator, i.e., those spatially independent fields for which $-\div\cdot\vec{D}\phi(x_0)=0$, one is left with \cite{Reinhardt:2008pr}
\be
Z=\!\!\int\!\!\!{\cal D}\Phi\de\!\left(\div\cdot\vec{A}\right)\!\de\!\left(\div\cdot\vec{\pi}\right)\!\de\!\left(\int d\vec{x}\rho\!\!\right)e^{\imath{\cal S}''},\;
{\cal S}''\sim\int\!\!dx\left[\ldots-\ro^a\hat{F}^{ab}\ro^b/2\right].
\ee
In the above, one sees that there are two transverse degrees of freedom for the gluon and the total color charge is conserved and vanishing.  The Coulomb kernel $\hat{F}=[-\div\cdot\vec{D}]^{-1}(-\nabla^2)[-\div\cdot\vec{D}]^{-1}$ is nonlocal in $\vec{A}$, so we make the leading order truncation whereby it is replaced by its expectation value, which is related to the temporal component of the gluon propagator: $\hat{F}\rightarrow\ev{\hat{F}}\sim W_{00}$.  It is known that in Coulomb gauge on the lattice, $W_{00}$ is infrared (IR) enhanced, going like $\si/\vec{q}^4$ but with a coefficient $\si$ somewhat larger than the Wilson string tension  (see e.g., Refs.~\cite{Iritani:2010mu,Zwanziger:2002sh}).  The charge conservation term is rewritten in the limiting form of a Gaussian, mimicking the Coulomb term:
\be
\de\left(\int d\vec{x}\rho\right)\sim\lim_{{\cal C}\rightarrow\infty}{\cal N}\exp{\left\{-\imath/2\int dx\,dy\ro_x^a\de^{ab}{\cal C}\de(x_0-y_0)\ro_y^b\right\}},
\ee
and such that we now have instantaneous four-point interaction terms including $\G_{AA\pi\pi}$ and $\G_{\ov{q}q\ov{q}q}$:
\be
{\cal S}_{\mbox{int}}\sim\int dx\,dy\,\left[-\rho_x^a\de^{ab}\tilde{F}_{xy}\rho_y^b/2\right],\;\;\;\;
g^2C_F\tilde{F}(\vec{q})=(2\pi)^3{\cal C}\de(\vec{q})+8\pi\si/\vec{q}^4
\ee
($C_F=(N_c^2-1)/2N_c$).  This interaction contains the charge constraint and leads directly to a linear rising potential with a string tension $\si$.  To complete the leading order truncation scheme, we restrict to one-loop terms in the following equations and disregard all but the $\tilde{F}$ interaction terms.

\section{Gluon gap equation}
\begin{figure}[t]
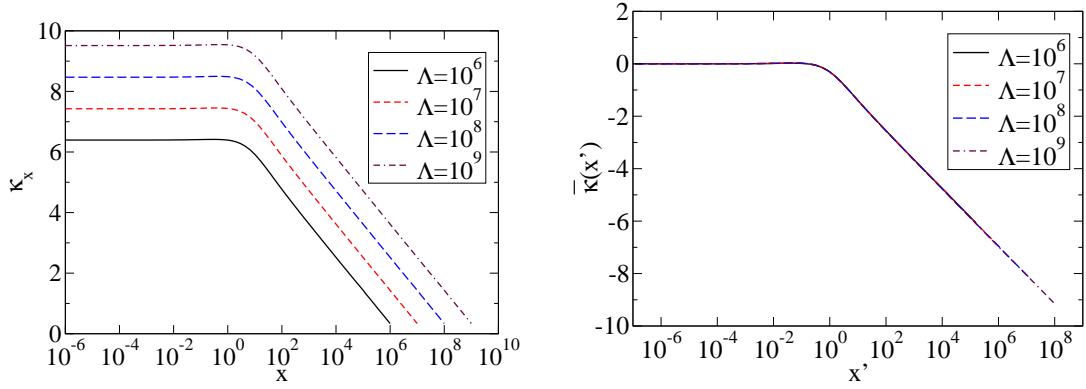

\vspace{0.8cm}
\begin{center}
\includegraphics[width=0.45\linewidth]{hap0.eps}
\hspace{0.5cm}
\includegraphics[width=0.45\linewidth]{hap1.eps}
\end{center}
\vspace{0.3cm}
\caption{\label{fig:hap0}[left panel] $\ka_x$ as a function of $x=\vec{k}^2$ and [right panel] $\ov{\ka}(x')$ as a function of $x'=x/a$ for various values of the UV-cutoff $\La$.  All dimensionfull quantities are in appropriate units of the string tension, $\si$.  See text for details.}
\end{figure}

In the first order formalism, the transverse spatial gluon degrees of freedom ($\vec{A}$, $\vec{\pi}$) have been separated such that there are three propagators $W_{AAij}$, $W_{\pi\pi ij}$ and $W_{A\pi ij}$ ($i,j$ are the spatial indices), correspondingly with three proper functions $\G_{AAij}$, $\G_{\pi\pi ij}$ and $\G_{A\pi ij}$ related via a matrix inversion structure (see, e.g., Ref.~\cite{Watson:2006yq}).  Since the interaction content of our truncated system is instantaneous, the energy dependence of these functions is trivial (and the mixed functions will play no role in the discussion here).  There are two scalar dressing functions of interest, both functions of spatial momentum: $\G_{AA}(\vec{k}^2)$ and $\G_{\pi\pi}(\vec{k}^2)$.  The spatial gluon propagator $W_{AA}$ has the form ($W_{\pi\pi}$ is similar)
\be
W_{AAij}(k)=\imath t_{ij}(\vec{k})\frac{\G_{\pi\pi}(\vec{k}^2)}{[k_0^2-\vec{k}^2\G_{AA}(\vec{k}^2)\G_{\pi\pi}(\vec{k}^2)+\imath0_+]}
\ee
($t_{ij}$ is the transverse spatial momentum projector).  The truncated \DS equations have the mnemonic form (omitting kinematical factors etc.)
\be
\G_{\pi\pi}(\vec{p}^2)\sim1+\int dk\tilde{F}(\vec{p}-\vec{k})W_{AA}(k),\;\;\;\;
\G_{AA}(\vec{p}^2)\sim1+\int dk\tilde{F}(\vec{p}-\vec{k})W_{\pi\pi}(k).
\ee
The charge constraint term of the interaction $\tilde{F}$ (i.e., the term $\sim{\cal C}\de(\vec{p}-\vec{k})$) immediately tells us that both $\G_{AA}$ and $\G_{\pi\pi}$ are divergent as ${\cal C}\rightarrow\infty$ (there is also an IR divergence), meaning that the gluon self-energy is infinite and the propagator poles are shifted to infinity.  This has the natural interpretation that one requires infinite energy to create a (colored) gluon from the (colorless) vacuum.  If, however, we consider the static gluon propagator $W_{AA}^{(s)}$, written as
\be
W_{AAij}^{(s)}(\vec{k})=\int\frac{dk_0}{2\pi}W_{AAij}(k)=t_{ij}(\vec{k})\frac{\sqrt{G_k}}{2|\vec{k}|},\;\;\;\;G_k=\frac{\G_{\pi\pi}(\vec{k}^2)}{\G_{AA}(\vec{k}^2)}
\ee
then we can combine the \DS equations to get the gluon gap equation
\be
G_p=1+\frac{g^2N_c}{4}\int\frac{d\vec{k}\,\tilde{F}(\vec{p}-\vec{k})}{(2\pi)^3|\vec{k}|}t_{ji}(\vec{p})t_{ij}(\vec{k})\left[\sqrt{G_k}-\frac{\vec{k}^2}{\vec{p\,}^2}\frac{G_p}{\sqrt{G_k}}\right].
\ee
This equation has previously been derived in the Coulomb gauge Hamiltonian approach \cite{Szczepaniak:2001rg}.  The dressing function for the static propagator, $G$, is IR finite and independent of the charge constraint.  Solving numerically (in units of $\si$), one sees that the solution has the form $G_x=x/(x+\ka_x)$ for an IR constant `mass' function $\ka(x)$ and where $x=\vec{k}^2$.  $\ka_x$ is logarithmically dependent on the numerical ultraviolet (UV) cutoff $\La$ (dimensions of $[\mbox{mass}]^2$), despite the fact that the interaction has the form $1/\vec{q}^4$.  $\ka$ is plotted in the left panel of Fig.~\ref{fig:hap0}.  However, defining $a=\ka(x=0)$ and introducing the scaled variable $x'=x/a$, one finds that $\ov{\ka}(x')=\ka(x=x'a)-a$ is independent of $\La$, shown in the right panel of Fig.~\ref{fig:hap0}.  It turns out in general that by simply writing all dimensionfull quantities in units of the (dynamically generated) gluon mass function at some point, one may construct $\La$-independent dressing functions (and subsequently $G$) without introducing a renormalization constant \cite{Watson:2011kv}.

\section{Quark gap equation}
\begin{figure}[t]
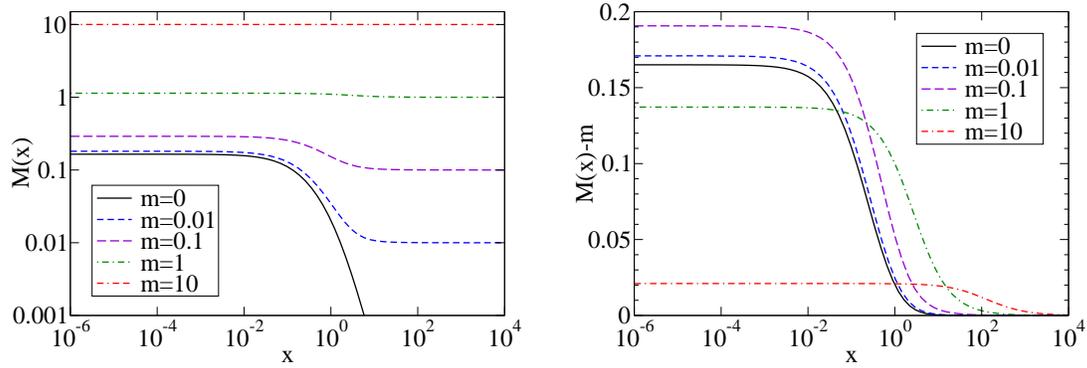

\vspace{0.8cm}
\begin{center}
\includegraphics[width=0.45\linewidth]{eap0.eps}
\hspace{0.5cm}
\includegraphics[width=0.45\linewidth]{eap1.eps}
\end{center}
\vspace{0.3cm}
\caption{\label{fig:mfunc}[left panel] Quark mass function, $M(x)$, and 
[right panel] dressing, $M(x)-m$, plotted as functions of $x=\vec{p\,}^2$ 
for a range of quark masses.  All dimensionfull quantities are in 
appropriate units of the string tension, $\si$.  See text for details.}
\end{figure}

Given that the interaction content arising from the Coulomb term couples to the gluon and quark charges in the same manner, the quark sector turns out to be very similar to the gluon sector within the leading order truncation scheme considered here.  The instantaneous character of the interaction leads immediately to the following form for the quark propagator in terms of two dressing functions $A$ and $B$ (both functions of $\vec{k\,}^2$):
\be
W_{\ov{q}q}(k)=(-\imath)\frac{\ga^0k_0-\vec{\ga}\cdot\vec{k}A_k+B_k}{[k_0^2-\vec{k}^2A_k^2-B_k^2+\imath0_+]}.
\ee
A possible term $\sim\ga^0k_0\vec{\ga}\cdot\vec{k}$ does not appear, just as in the perturbative case \cite{Popovici:2008ty}.  The static quark propagator, $W_{\ov{q}q}^{(s)}$, can be written in terms of the mass function, $M$, and quasiparticle energy, $\w$:
\be
W_{\ov{q}q}^{(s)}(\vec{k})=\int\frac{dk_0}{2\pi}W_{\ov{q}q}(\vec{k})=\frac{\vec{\ga}\cdot\vec{k}-M_k}{2\w_k},\;\;\;\;M_k=\frac{B_k}{A_k},\;\;\w_k^2=\vec{k}^2+M_k^2.
\ee
The \DS equations for the dressing functions $A$ and $B$ have the mnemonic form
\be
A_p\sim1+\int d\vec{k}\tilde{F}(\vec{p}-\vec{k})/\w_k,\;\;\;\;B_p\sim m+\int d\vec{k}\tilde{F}(\vec{p}-\vec{k})M_k/\w_k,
\ee
showing us via the charge constraint that the quark self-energy is divergent (like for the gluon) and one requires infinite energy to extract a single quark from the vacuum.  However, combining the equations in terms of the mass function, $M$, leads to the Adler-Davis gap equation \cite{Adler:1984ri}
\be
M_p=m+\frac{1}{2}g^2C_F\int\frac{d\vec{k}\,\tilde{F}(\vec{p}-\vec{k})}{(2\pi)^3\w_k}\left[M_k-\frac{\vec{p}\cdot\vec{k}}{\vec{p\,}^2}M_p\right].
\ee
The mass function is IR finite and independent of the charge constraint.  While the above equation was originally derived for chiral quarks (in the Hamiltonian formalism), in the leading order truncation scheme presented here one can show \cite{Watson:2011kv} that it also reproduces the Coulomb gauge heavy quark limit (in the absence of pure Yang-Mills corrections) \cite{Popovici:2010mb}.  The mass function is plotted on the left panel of Fig.~\ref{fig:mfunc}.  One can see that chiral symmetry is dynamically broken, although the chiral condensate is too small \cite{Watson:2012ht} (this can be improved by considering the spatial quark-gluon vertex \cite{Pak:2011wu}).  In the right panel of Fig.~\ref{fig:mfunc}, the dressing $M(x)-m$ is plotted.  As the quark mass increases the dressing initially also increases, but for heavier quarks becomes smaller and in the heavy quark limit, $M\rightarrow m$.

\section{Bethe-Salpeter equation}
Within this leading order truncation scheme, it is possible to study the quark-antiquark \BS equation for (color singlet, flavor nonsinglet) pseudoscalar and vector mesons with arbitrary quark masses \cite{Watson:2012ht}.  The pseudoscalar case will be discussed here -- the vector case is similar.  In the Coulomb gauge rest frame, the \BS vertex for pseudoscalar meson can be written (omitting flavor factors)
\be
\G_{PS}(\vec{p};P_0)=\ga^5\left[\G_0+P_0\ga^0\G_1+\vec{\ga}\cdot\vec{p}\,\G_2+P_0\ga^0\vec{\ga}\cdot\vec{p}\,\G_3\right],
\ee
where $P_0^2=M_{PS}^2$ is the total energy squared (at resonance) for the quark-antiquark pair and $\vec{p}$ the spatial momentum flowing along the quark line.  The dressing functions $\G_i$ all have the argument $\vec{p\,}^2$.  There are two basic quantities of initial interest (trace over Dirac matrices):
\be
\left\{\begin{array}{c}f_{PS}\\h_{PS}\end{array}\right\}=\frac{N_c}{M_{PS}^2}\mbox{Tr}_d\int\frac{dk}{(2\pi)^4}
\left\{\begin{array}{c}\ga^5P_0\ga^0\\M_{PS}^2\ga^5\end{array}\right\}
W_{\ov{q}q}^+(k^+)\G_{PS}(\vec{k};P_0)W_{\ov{q}q}^-(k^-),
\label{eq:fhps}
\ee
where $k^\pm$ represents the energy and spatial momentum argument $(k^0\pm P_0/2,\vec{k})$ and the two quark propagators $W_{\ov{q}q}^\pm$ correspond to bare quark masses $m^\pm$.  $f_{PS}$ is the pseudoscalar meson leptonic decay constant.  $h_{PS}$ is related to $f_{PS}$ via the axialvector Ward-Takahashi identity (AXWTI) \cite{Watson:2012ht,Maris:1997hd} and this can be compared to the Gell-Mann-Oakes-Renner relation in the chiral limit
\be
M_{PS}^2f_{PS}=(m^++m^-)h_{PS},\;\;\;\;h_{PS}\stackrel{m^\pm\rightarrow0}{\longrightarrow}-\ev{\ov{q}q}/f_{PS},
\ee
indicating that $h_{PS}$ is a generalization of the chiral condensate to finite, arbitrary mass quarks.  Evaluating the trace and energy integrals for the right-hand side of \eq{eq:fhps}, one obtains spatial integrals over a combination of terms involving IR divergent quantities such as $A$.  However, $f_{PS}$ and $h_{PS}$ must be IR finite.  Assuming the form
\be
f_{PS}=2\imath N_c\int\frac{d\vec{k}}{(2\pi)^3\w_k^+\w_k^-}\frac{[M_k^++M_k^-]}{[\w_k^++\w_k^-]}f_k,\;\;h_{PS}=2\imath N_c\int\frac{d\vec{k}}{(2\pi)^3\w_k^+\w_k^-}[\w_k^++\w_k^-]h_k,
\ee
the combinations of divergent factors are then contained within the two functions $f$ and $h$.  Here is where Coulomb gauge does something special: when one expands the truncated \BS equation
\be
\G_{PS}(\vec{p}\,;P^0)=-\imath g^2C_F\int\frac{dk}{(2\pi)^4}\tilde{F}(\vec{p}-\vec{k})\ga^0W_{\ov{q}q}^+(k^+)\G_{PS}(\vec{k};P_0)W_{\ov{q}q}^-(k^-)\ga^0,
\ee
the right-hand side takes the mnemonic form $\G_i\sim\int\tilde{F}[\ldots][f_k\;\mbox{or}\;h_k]$ where the terms represented by the dots ($[\dots]$) involve combinations of only the finite functions $M_k^\pm$ or $\w_k^\pm$.  The \BS equation can thus be rewritten in terms of only $f$ and $h$.  The equal mass case is
\begin{align}
h_p&=\frac{P_0^2}{4\w_p^2}f_p
+\frac{1}{2}g^2C_F\int\frac{d\vec{k}\tilde{F}(\vec{p}-\vec{k})}{(2\pi)^3\w_k}
\left\{h_k-h_p\frac{\s{\vec{p}}{\vec{k}}}{\vec{p\,}^2}\right\},
\nonumber\\
f_p&=h_p
+\frac{1}{2}g^2C_F\int\frac{d\vec{k}\tilde{F}(\vec{p}-\vec{k})}{(2\pi)^3\w_k}
\left\{
f_k\frac{\left[\s{\vec{p}}{\vec{k}}+M_pM_k\right]}
{\left[\vec{k}^2+M_k^2\right]}
-f_p\frac{\s{\vec{p}}{\vec{k}}}{\vec{p\,}^2}
\right\}.
\label{eq:psbsee}
\end{align}
The arbitrary mass case has a similar form.  The corresponding vector meson equation is also similar, but involves four functions.  One can see that the above form for the \BS equation thus behaves like the previously discussed gap equations for $G$ and $M$, where the charge constraint and IR divergences cancel and despite the interaction, the functions $f$ and $h$ are finite.  The equations can be compared to those of, for example, Refs.\cite{Govaerts:1983ft}.

\begin{figure}[t]
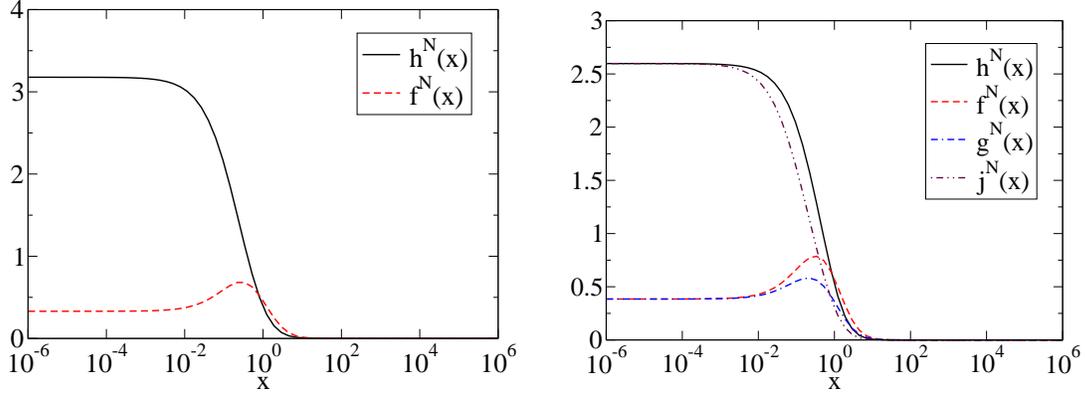

\vspace{0.8cm}
\begin{center}
\includegraphics[width=0.45\linewidth]{reseqvertps.eps}
\hspace{0.5cm}
\includegraphics[width=0.45\linewidth]{reseqvertv.eps}
\end{center}
\vspace{0.3cm}
\caption{\label{fig:eqvert}[left panel] Pseudoscalar and [right panel] 
vector normalized vertex functions with (equal) chiral quarks, plotted 
as a function of $x=\vec{k}^2$.  All dimensionfull quantities are in 
appropriate units of the string tension, $\si$.  See text for details.}
\end{figure}

\begin{figure}[t]
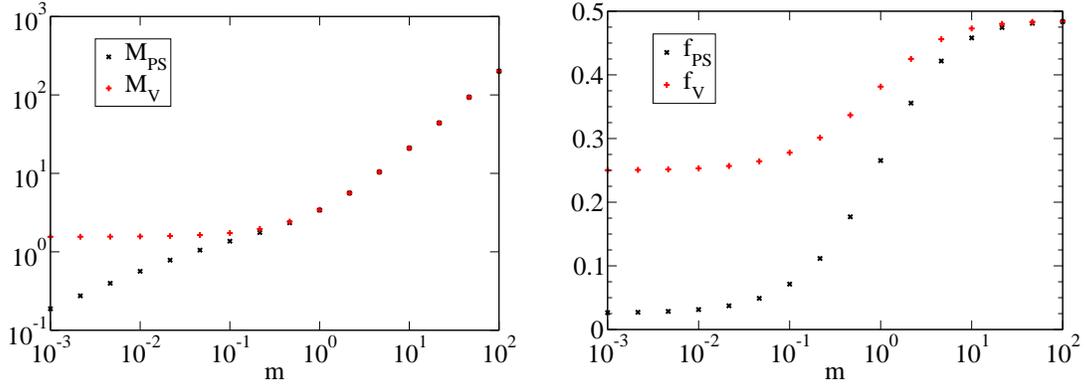

\vspace{0.8cm}
\begin{center}
\includegraphics[width=0.45\linewidth]{reseqml.eps}
\hspace{0.5cm}
\includegraphics[width=0.45\linewidth]{reseqfl.eps}
\end{center}
\vspace{0.3cm}
\caption{\label{fig:eqm}Pseudoscalar and vector meson masses 
[left panel] and leptonic decay constants [right panel] with equal mass quarks, 
plotted as a function of the quark mass.  All dimensionfull quantities 
are in appropriate units of the string tension, $\si$.  See text for 
details.}
\end{figure}

\begin{figure}[t]
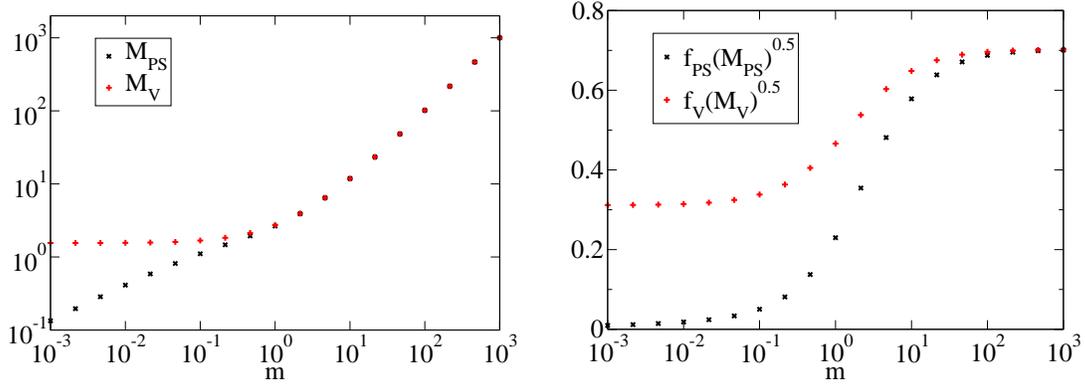

\vspace{0.5cm}
\begin{center}
\includegraphics[width=0.45\linewidth]{reshlm.eps}
\hspace{0.5cm}
\includegraphics[width=0.45\linewidth]{reshlfm.eps}
\end{center}
\vspace{0.0cm}
\caption{\label{fig:hlm}Pseudoscalar and vector meson masses 
[left panel] and $f_{PS}\sqrt{M_{PS}}$, $f_V\sqrt{M_V}$ [right panel] with one fixed chiral 
quark, plotted as a function of the other quark mass.  All dimensionfull 
quantities are in appropriate units of the string tension, $\si$.  
See text for details.}
\end{figure}

Turning to the results, the normalized (see Ref.~\cite{Watson:2012ht}) pseudoscalar and vector meson dressing functions are plotted for the chiral quark case in Fig.~\ref{fig:eqvert} and one sees that indeed, the functions are all finite.
In Fig.\ref{fig:eqm}, the meson masses and leptonic decay constants for equal quark masses are plotted (in units of the string tension $\si$).  Inserting typical values for $\si$ \cite{Watson:2012ht}, it becomes obvious that whilst dynamical chiral symmetry breaking is visible ($M_{PS}\sim\sqrt{m}$ as $m\rightarrow0$), the leptonic decay constants are too small, as is the mass-splitting between states for larger quark masses.
Looking at the case for one fixed chiral quark plotted in Fig.~\ref{fig:hlm}, one sees that both the pattern for chiral symmetry breaking ($M_{PS}\sim\sqrt{m}$ as $m\rightarrow0$) and the leading order heavy quark limit ($f_{PS}\sqrt{M_{PS}},f_{V}\sqrt{M_{V}}\sim\mbox{const.}$ as $m\rightarrow\infty$) are present.  The leading order Coulomb gauge truncation scheme thus qualitatively accommodates both chiral and heavy quark physics.

\begin{acknowledgments}
It is a pleasure to thank the organizers for a most enjoyable and stimulating conference.
\end{acknowledgments}


\begin{thebibliography}{99}
\bibitem{Watson:2011kv} 
  P.~Watson and H.~Reinhardt,
  Phys.\ Rev.\ D {\bf 85}, 025014 (2012)
  [arXiv:1111.6078 [hep-ph]].

\bibitem{Reinhardt:2008pr} 
  H.~Reinhardt and P.~Watson,
  Phys.\ Rev.\ D {\bf 79}, 045013 (2009)
  [arXiv:0808.2436 [hep-th]].

\bibitem{Iritani:2010mu} 
  T.~Iritani and H.~Suganuma,
  Phys.\ Rev.\ D {\bf 83}, 054502 (2011)
  [arXiv:1011.4767 [hep-lat], arXiv:1102.0920 [hep-lat]].

\bibitem{Zwanziger:2002sh} 
  D.~Zwanziger,
  Phys.\ Rev.\ Lett.\  {\bf 90}, 102001 (2003)
  [hep-lat/0209105].

\bibitem{Watson:2006yq} 
  P.~Watson and H.~Reinhardt,
  Phys.\ Rev.\ D {\bf 75}, 045021 (2007)
  [hep-th/0612114];
  Phys.\ Rev.\ D {\bf 76}, 125016 (2007)
  [arXiv:0709.0140 [hep-th]].

\bibitem{Szczepaniak:2001rg} 
  A.~P.~Szczepaniak and E.~S.~Swanson,
  Phys.\ Rev.\ D {\bf 65}, 025012 (2002)
  [hep-ph/0107078];
  C.~Feuchter and H.~Reinhardt,
  Phys.\ Rev.\ D {\bf 70}, 105021 (2004)
  [hep-th/0408236];
  hep-th/0402106.

\bibitem{Popovici:2008ty} 
  C.~Popovici, P.~Watson and H.~Reinhardt,
  Phys.\ Rev.\ D {\bf 79}, 045006 (2009)
  [arXiv:0810.4887 [hep-th]].

\bibitem{Adler:1984ri} 
  S.~L.~Adler and A.~C.~Davis,
  Nucl.\ Phys.\ B {\bf 244}, 469 (1984).

\bibitem{Popovici:2010mb} 
  C.~Popovici, P.~Watson and H.~Reinhardt,
  Phys.\ Rev.\ D {\bf 81}, 105011 (2010)
  [arXiv:1003.3863 [hep-th]];
  Phys.\ Rev.\ D {\bf 83}, 025013 (2011)
  [arXiv:1010.4254 [hep-ph]];
  Phys.\ Rev.\ D {\bf 83}, 125018 (2011)
  [arXiv:1103.4786 [hep-ph]].

\bibitem{Watson:2012ht} 
  P.~Watson and H.~Reinhardt,
  arXiv:1211.4507 [hep-ph].

\bibitem{Pak:2011wu} 
  M.~Pak and H.~Reinhardt,
  Phys.\ Lett.\ B {\bf 707}, 566 (2012)
  [arXiv:1107.5263 [hep-ph]].

\bibitem{Maris:1997hd} 
  P.~Maris, C.~D.~Roberts and P.~C.~Tandy,
  Phys.\ Lett.\ B {\bf 420}, 267 (1998)
  [nucl-th/9707003].

\bibitem{Govaerts:1983ft} 
  J.~Govaerts, J.~E.~Mandula and J.~Weyers,
  Nucl.\ Phys.\ B {\bf 237}, 59 (1984);
  R.~Alkofer and P.~A.~Amundsen,
  Nucl.\ Phys.\ B {\bf 306}, 305 (1988);
  R.~F.~Wagenbrunn and L.~Y.~.Glozman,
  Phys.\ Rev.\ D {\bf 75}, 036007 (2007)
  [hep-ph/0701039].


\end{thebibliography}
\end{document}